\pdfoutput=1
\documentclass[acmsmall,screen,nonacm]{acmart}

\setcopyright{acmlicensed}
\copyrightyear{2027}
\acmYear{2027}
\acmDOI{XXXXXXX.XXXXXXX}
\acmJournal{PACMSE}
\acmVolume{4}
\acmNumber{FSE}
\acmMonth{7}

\newcommand{\tool}[1]{\textsc{#1}}
\newcommand{\finding}[1]{\par\indent\textbf{Finding.}~#1\par}

\usepackage{algorithm}
\usepackage{algorithmic}
\usepackage{subcaption}
\usepackage{tikz}
\usepackage{tabularx}
\usepackage{multirow}
\usepackage{quantikz}

\usepackage{amsmath}
\usepackage{booktabs}
\usepackage{array}
\usepackage{caption}

\DeclareCaptionFont{tablefont}{\fontsize{6}{7}\selectfont}
\DeclareCaptionFont{tablelabel}{\fontsize{7}{8}\selectfont}
\DeclareCaptionLabelFormat{spacedtable}{#1\hspace{0.4em}#2}
\DeclareCaptionLabelSeparator{spacedperiod}{.\hspace{0.6em}}

\begin{document}

\title{Isolating Recurring Backend-Dependent Abnormal Patterns on NISQ Quantum Devices}

\author{Zhenyu Qi}
\affiliation{%
  \institution{University of Arizona}
  \city{Tucson}
  \state{AZ}
  \country{USA}}
\email{qzydustin@arizona.edu}

\author{Haotang Li}
\affiliation{%
  \institution{University of Arizona}
  \city{Tucson}
  \state{AZ}
  \country{USA}}
\email{haotangl@arizona.edu}

\author{Mominul Islam}
\affiliation{%
  \institution{Tulane University}
  \city{New Orleans}
  \state{LA}
  \country{USA}}
\email{mislam9@tulane.edu}

\author{Qian Zhang}
\affiliation{%
  \institution{University of California, Riverside}
  \city{Riverside}
  \state{CA}
  \country{USA}}
\email{qzhang@cs.ucr.edu}

\author{Sen He}
\affiliation{%
  \institution{University of Arizona}
  \city{Tucson}
  \state{AZ}
  \country{USA}}
\email{senhe@arizona.edu}

\author{Jiyuan Wang}
\authornote{Corresponding author.}
\affiliation{%
  \institution{Tulane University}
  \city{New Orleans}
  \state{LA}
  \country{USA}}
\email{wjiyuan@tulane.edu}

\renewcommand{\shortauthors}{Qi et al.}

\begin{abstract}
Quantum hardware remains highly susceptible to noise, making compilation a critical step for obtaining reliable execution on existing devices. Today’s quantum compilers approximate hardware noise using calibration-derived costs over gates, qubits, and coupling edges, and use the costs to guide qubit mapping, routing, and scheduling. However, this abstraction is incomplete: semantically equivalent circuits with identical compiler-predicted hardware noise can still behave very differently on real devices. Execution context can contribute to this gap: short ordered gate sequences on specific qubit neighborhoods can be associated with hardware--model mismatch.
Identifying these backend-dependent abnormal patterns can guide gate-ordering rewrites in the circuit family and layout where they were observed.

To find these patterns, we propose \tool{QRisk}, a framework that isolates recurring abnormal gate sequences from quantum hardware executions. \tool{QRisk} starts from a circuit whose hardware output is much worse than predicted by the calibration-based noise model. It then reduces the circuit at the segment level, identifying groups of segments whose removal lowers the discrepancy ratio or whose retention within the tested circuit context preserves an elevated ratio. The resulting circuit fragments are treated as candidate abnormal patterns.

On three IBM Heron r2 backends (quantum devices) across 30 qubit layouts, \tool{QRisk} finds 25 recurring abnormal gate patterns.
Reordering commuting gates to break the two highest-recurrence patterns in their Grover discovery circuits reduces mean hardware--model mismatch by 24\% on \texttt{ibm\_fez} and 45\% on \texttt{ibm\_marrakesh}.
\end{abstract}




\maketitle

\section{Introduction}
\label{sec:intro}

Quantum compilers translate logical circuits into hardware-executable circuits by mapping logical qubits to physical qubits, routing two-qubit interactions through the device's coupling graph, and decomposing gates into native operations~\cite{hua2023qasmtrans,javadiabhari2024qiskit}. On noisy intermediate-scale quantum (NISQ) devices~\cite{preskill2018nisq}, compilation decisions affect execution fidelity, and compilers use calibration data and noise model to guide these choices.

Existing noise-aware compilation strategies address error sources that can be characterized statically~\cite{murali2020software,zhu2025compiler}. A method may use per-gate error rates for layout scoring, crosstalk characterization data for scheduling constraints, or decoherence times for idle-time management, but in each case the error source is known in advance and encoded into the compiler through calibration measurements. Section~\ref{sec:noise_aware} details what current compilers use and what they ignore.
However, static compilation cannot detect effects that emerge only during execution.
Prior work shows that quantum-processor behavior can depend on execution context, including spectator-qubit activity and time~\cite{rudinger2019probing}.
We study recurring \emph{backend-dependent abnormal patterns}: short gate sequences on specific physical qubit neighborhoods associated with hardware--model mismatch across observation windows.
\textbf{Production noise-aware compilers today do not learn such backend-dependent abnormal patterns from hardware execution.}
We use total variation distance (TVD) to measure differences between output distributions, with smaller values indicating closer agreement.

\begin{figure*}[!t]
  \centering
  \includegraphics[width=0.95\textwidth]{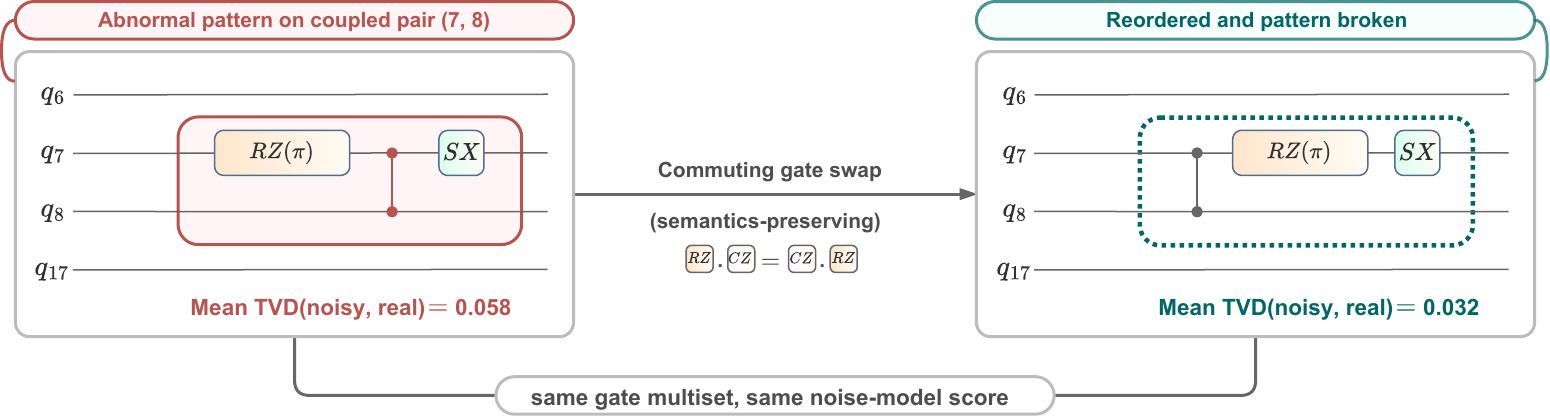}
  \caption{An example on \texttt{ibm\_marrakesh} (qubits 6, 7, 8, 17). Two compiler-equivalent circuits (same gate multiset, same noise-model score) produce different hardware error. The high-error version contains a 3-gate pattern. A commuting gate swap breaks the pattern without changing semantics. Mean TVD over 10 variants per level falls from 0.058 (3 occurrences) to 0.032 (0 occurrences), a 45\% reduction in the controlled intervention.}
  \Description{Motivating example on ibm_marrakesh showing two compiler-equivalent circuits with different hardware error. The difference is a three-gate pattern that a commuting swap can break.}
  \label{fig:motivating}
\end{figure*}

Figure~\ref{fig:motivating} shows an example on \texttt{ibm\_marrakesh}. A compiled Grover circuit contains a 3-gate sequence on two coupled qubits. The sequence recurs in 10 of 18 observation windows. A single commuting gate swap breaks the contiguous pattern. In our experiment (Section~\ref{sec:eval-rq3}), reducing the number of pattern occurrences from three to zero through such swaps correlates with a 45\% reduction in mean hardware--model mismatch, even though the calibration-based noise model assigns identical scores to all variants.

\begin{figure}[!t]
  \centering
  \includegraphics[width=0.85\textwidth]{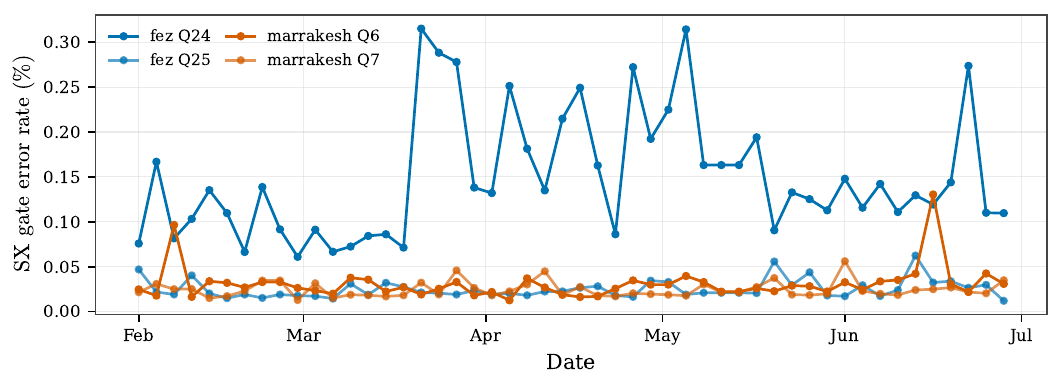}
  \caption{Historical gate error rate drift on \texttt{ibm\_fez} and \texttt{ibm\_marrakesh} (February--July 2026). SX gate error rates for selected qubits sampled every 3 days show fluctuation and drift between recalibration events, demonstrating hardware instability that necessitates cross-window validation.}

  \Description{Time series showing SX gate error rates for two qubits on each of two IBM backends over five months, exhibiting drift patterns from real hardware calibration data sampled every 3 days.}
  \label{fig:calibration-drift}
\end{figure}

To identify these backend-dependent abnormal patterns, we propose \tool{QRisk}, a framework that leverages delta debugging~\cite{zeller2002simplifying} to isolate candidate gate sequences and validate their recurrence across calibration windows, observation periods with separate calibration snapshots.
Prior work applies delta debugging to property-based regression testing of quantum programs~\cite{pontolillo2024delta} and statistical testing to locate buggy quantum-program segments~\cite{sato2024locating}.
These methods target program regressions or correctness bugs rather than recurring hardware--model mismatches under calibration drift.
Adapting delta debugging to this setting presents three challenges:

\begin{itemize}
\item \textit{Circuit reduction changes the expected error.}
Removing gates can lower ordinary error even when none of them belongs to an abnormal pattern.
A naive pass/fail oracle, a decision rule for evaluating circuit reductions, therefore cannot tell whether the observed error reduction comes from removing an abnormal pattern or simply from shortening the circuit.
The oracle must compare observed hardware error with the error predicted for the same reduced circuit, accounting for the change in expected error as gates are removed.

\item \textit{Finite sampling makes the failure signal stochastic.}
Each shot is one circuit execution and measurement; repeated shots yield a sampled output distribution.
The oracle must distinguish a meaningful discrepancy from sampling fluctuations.

\item \textit{Hardware behavior drifts over time}~\cite{klimov2018fluctuations,ravi2023qismet}.
IBM backends are recalibrated periodically, and error rates shift between calibration windows, as shown in Figure~\ref{fig:calibration-drift}.
A fragment that appears abnormal in one window may therefore reflect a temporary fluctuation rather than a recurring effect that can guide later compilations.
\end{itemize}

To tackle these challenges, \tool{QRisk} uses a ratio-based stochastic oracle to guide circuit reduction and cross-window validation to identify persistent patterns.
The oracle normalizes observed hardware error by the Aer reference model's prediction for each reduced circuit and uses calibrated thresholds to distinguish error changes from sampling fluctuations.
Only candidates independently rediscovered across calibration windows enter the backend-specific pattern database, providing compilation guidance based on repeated observations.

We evaluate \tool{QRisk} on three IBM Heron r2 backends across 30 qubit layouts covering 117 backend-specific physical qubits.
It isolates 3--6-gate candidate patterns from circuits of 151--158 operations, and cross-window validation retains 25 of the 92 distinct candidates.
The controlled intervention reduces mean hardware--model mismatch by 24\% on \texttt{ibm\_fez} and 45\% on \texttt{ibm\_marrakesh}.
Additional experiments on MQT~Bench workloads extend isolation to amplitude estimation, quantum walk, and random circuits, yielding compact fragments in 11--15 oracle calls.
The cross-backend comparison supports validating patterns on their target hardware, even when two devices share the same topology (Section~\ref{sec:cross-backend}).

This paper makes three contributions:
\begin{itemize}
    \item We identify and empirically validate backend-dependent abnormal patterns: ordered gate sequences on specific physical qubits associated with hardware--model mismatch in multiple calibration windows.
    \item We develop \tool{QRisk}, which combines delta debugging, a stochastic discrepancy oracle, and cross-window validation to isolate these patterns and build a backend-specific database.
    \item We evaluate pattern isolation and persistence across backends and layouts, demonstrate compilation improvements through controlled gate swaps, and assess isolation on additional benchmark workloads.
\end{itemize}

\section{Background}
\label{sec:background}

This section introduces the technical foundations of \tool{QRisk}.
Section~\ref{sec:qc-compilation} describes quantum circuits and compilation, and Section~\ref{sec:noise_aware} reviews the calibration data and noise model used by static error-aware compilers.
Section~\ref{sec:dd-background} introduces delta debugging, while Section~\ref{sec:mismatch-signal} explains how hardware error is compared with noise-model predictions.

\subsection{Quantum Circuits and Compilation}
\label{sec:qc-compilation}

A quantum computation operates on \emph{qubits}. Unlike a classical bit, a qubit can exist in a superposition of $|0\rangle$ and $|1\rangle$; measurement collapses the superposition and returns a classical outcome. Computations are expressed as \emph{quantum circuits}: sequences of \emph{quantum gates} applied to qubits. Each gate corresponds to a unitary matrix, and a circuit implements the matrix product of its gates. Two circuits are \emph{semantically equivalent} if they produce the same measurement statistics.

A \emph{backend} is the quantum device on which a circuit executes.
Physical qubits have limited connectivity described by a \emph{coupling graph}. A quantum compiler maps a logical circuit onto hardware through: (1)~\emph{qubit mapping}, assigning logical qubits to physical qubits; (2)~\emph{gate routing}, inserting SWAP operations so two-qubit gates act on connected pairs; and (3)~\emph{gate decomposition}, rewriting gates into the device's native gate set~\cite{hua2023qasmtrans,zhu2025compiler}. Two semantically equivalent compiled circuits may use different physical qubits, gate orderings, and SWAP sequences, producing different hardware error.

Two adjacent gates commute when their unitaries satisfy $U_1 U_2 = U_2 U_1$, so exchanging them does not change the circuit's semantics. For example, \texttt{rz} and \texttt{cz} are both diagonal in the computational basis and can be reordered on their shared qubit, while \texttt{sx} and \texttt{cz} on the same qubit cannot. Compilers use commutation rules for gate cancellation and circuit-mapping optimization~\cite{itoko2020optimization}. A \emph{commuting-gate swap} reorders one adjacent commuting pair, which changes the gate order but keeps the gate multiset and the circuit semantics.

\subsection{Static Error-Aware Compilation}
\label{sec:noise_aware}

Every gate and measurement on NISQ hardware is imperfect. Errors arise from imprecise control pulses, environmental decoherence ($T_1$/$T_2$ decay), and interactions between neighboring qubits~\cite{preskill2018nisq}. Compilers use \emph{calibration data} (periodically measured device characteristics) to guide compilation toward lower-error configurations. We use the term \emph{static error-aware compilation} to denote approaches driven by error signals characterized before circuit execution.

\indent \textit{Use of calibration data.} Qiskit's optimization level~3 (Qiskit 1.3.1, used in RQ1--RQ3) uses backend calibration as follows~\cite{qiskit13transpiler,javadiabhari2024qiskit}. \emph{Layout selection} (VF2) scores candidate mappings as $\sum_g -\log(1 - e_g)$, where $e_g$ is a scalar error rate for gate $g$, and selects the lowest-scoring layout. \emph{Routing} (SABRE) selects SWAP insertions by minimizing physical distances; error rates are not consulted~\cite{li2019sabre}. \emph{Peephole synthesis} resynthesizes two-qubit blocks and selects the variant with (1) fewest two-qubit gates, or if tied, (2) highest fidelity product. \emph{Identity removal} removes gates with unitaries close to identity.

\indent \textit{Limits of calibration data.} The calibration data available to the transpiler consists of a scalar \texttt{gate\_error} and \texttt{gate\_length} per gate per qubit (or qubit pair), plus per-qubit $T_1$, $T_2$, and frequency values. While $T_1$/$T_2$ are stored in the backend target, no optimization pass uses them for gate-level decisions. This compilation workflow does not use past hardware executions to identify recurring gate-sequence effects.

\subsection[Delta Debugging]{Delta Debugging}
\label{sec:dd-background}

Delta debugging~\cite{zeller2002simplifying} automatically reduces a failure-inducing input to a small subset that still causes the failure. Its core algorithm, DDMin, splits the input into $n$ parts and tests each part and each complement against an \emph{oracle} that reports whether the failure still occurs. When a smaller input still fails, DDMin continues from it with $n=2$, and otherwise it doubles $n$. It terminates with a \emph{1-minimal} input, from which removing any single part makes the failure disappear, using $O(m^2)$ oracle calls in the worst case for an input of $m$ parts. DDMin assumes a deterministic oracle: the same input always passes or always fails. On quantum hardware the oracle compares sampled output distributions against a model that drifts between calibrations, so this assumption does not hold. Section~\ref{sec:methodology} adapts DDMin to this setting.

\subsection{Hardware--Model Mismatch}
\label{sec:mismatch-signal}
Two models are relevant in this paper. The \emph{compiler cost model} uses calibration-derived scalar error rates to score layout and synthesis choices. The \emph{Aer reference noise model} constructs a per-gate depolarizing/thermal channel from the same calibration data and produces a simulated output distribution for any compiled circuit~\cite{qiskitaer2025noisemodel}. Neither model contains gate-sequence interaction terms.

Semantically equivalent circuits can share the same physical qubits, gate multiset, and layout score (Figure~\ref{fig:background-discrepancy}).
If the Aer reference model predicts similar output distributions for these circuits but their hardware outputs differ, the discrepancy indicates behavior that the reference model does not capture.
\tool{QRisk} uses this \emph{hardware--model mismatch} as a candidate failure signal for isolation.

We quantify differences between output distributions using total variation distance, $\mathrm{TVD}(P,Q)=\tfrac{1}{2}\sum_x |P(x)-Q(x)|$.
TVD ranges from 0 for identical distributions to 1 for distributions with no overlapping probability mass.

\begin{figure}[!t]
\centering
\begingroup
\newcommand{\outputbars}[5]{%
\begin{tikzpicture}[x=0.44cm,y=0.75cm,baseline]
\draw[gray!60] (-0.3,0)--(3.5,0);
\foreach \h [count=\i from 0] in {#2,#3,#4,#5} {
\fill[#1] (\i-0.25,0) rectangle (\i+0.25,\h);
}
\foreach \i/\lab in {0/00,1/01,2/10,3/11}
\node[font=\scriptsize,anchor=north] at (\i,-0.08) {\lab};
\end{tikzpicture}}
\begin{tabular}{@{}ccc@{}}
\textbf{Equivalent circuits} & \textbf{Aer predictions} & \textbf{Hardware outputs} \\
\begin{quantikz}[column sep=0.18cm,row sep=0.07cm]
\lstick{$q_0$} & \gate{R_z(\theta)} & \ctrl{1} & \qw \\
\lstick{$q_1$} & \qw & \ctrl{-1} & \qw
\end{quantikz}
& \outputbars{blue!60}{1.1}{0.3}{0.3}{0.3}
& \outputbars{orange!75}{0.5}{0.5}{0.5}{0.5} \\[6pt]
\begin{quantikz}[column sep=0.18cm,row sep=0.07cm]
\lstick{$q_0$} & \ctrl{1} & \gate{R_z(\theta)} & \qw \\
\lstick{$q_1$} & \ctrl{-1} & \qw & \qw
\end{quantikz}
& \outputbars{blue!60}{1.1}{0.3}{0.3}{0.3}
& \outputbars{orange!75}{0.9}{0.4}{0.4}{0.3} \\
\end{tabular}
\endgroup
\caption{Schematic of hardware--model mismatch. Reordering commuting gates preserves semantics, physical qubits, and the gate multiset. Similar Aer predictions can accompany different hardware outputs. The circuits represent local blocks within a larger circuit.}
\Description{Two local circuit blocks differ only in the order of commuting RZ and CZ gates. Blue bars illustrate matching model predictions, while orange bars illustrate differing hardware output distributions.}
\label{fig:background-discrepancy}
\end{figure}

Classical fault isolation reduces a failing input to a minimal subset that still triggers the failure.
\tool{QRisk} uses hardware--model mismatch to guide segment-level reduction and identify candidate abnormal gate sequences.

\section{QRisk Methodology}
\label{sec:methodology}

This section presents the design of \tool{QRisk}.
Section~\ref{sec:overview} gives a pipeline overview, and Sections~\ref{sec:oracle} and~\ref{sec:ddmin} describe the stochastic discrepancy oracle and circuit reduction procedure.
Section~\ref{sec:validation} describes cross-window recurrence validation.

\subsection{Overview}
\label{sec:overview}

\begin{figure*}[!t]
  \centering
  \includegraphics[width=\textwidth]{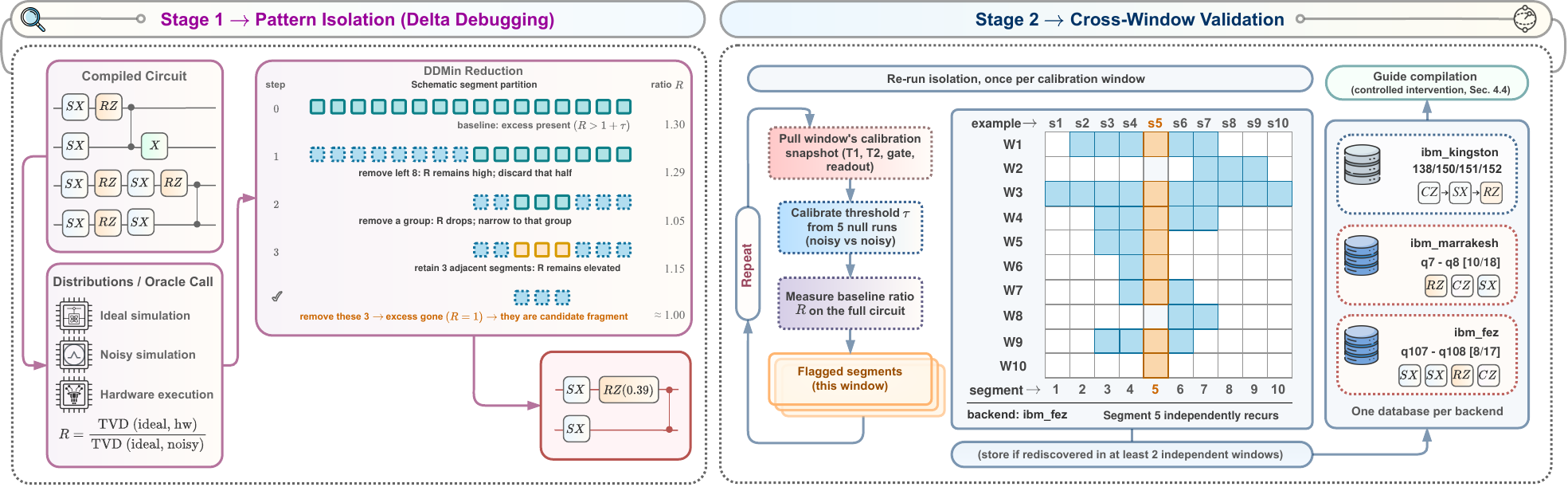}
  \caption{\tool{QRisk}'s two-stage workflow.
  DDMin isolates candidate segments, and independent rediscovery across calibration windows determines which patterns enter the backend-specific database.
  The segment partition, reduction ratios, and window grid are illustrative; database recurrence counts summarize the full observation campaigns (Tables~\ref{tab:inventory} and~\ref{tab:cost}).}
  \Description{Two-stage workflow: candidate isolation with a discrepancy-ratio oracle, followed by independent rediscovery across calibration windows and storage in a backend-specific pattern database. Reduction tests and the window grid are schematic.}
  \label{fig:overview}
\end{figure*}

Figure~\ref{fig:overview} shows the \tool{QRisk} workflow. \tool{QRisk} starts from circuits with known output distributions and uses them to expose gaps between noisy-simulator predictions and real hardware execution. Similar to traditional delta debugging, \tool{QRisk} first partitions the input circuit into moment-based segments. It then applies delta-debugging-based reduction with a stochastic discrepancy oracle to isolate compact candidate fragments associated with hardware--model mismatch, as described in Sections~\ref{sec:oracle} and~\ref{sec:ddmin}. After obtaining candidate patterns, \tool{QRisk} performs cross-window validation: it retains only patterns independently rediscovered across multiple calibration windows, as detailed in Section~\ref{sec:validation}. Verified patterns are stored in a backend-specific pattern database.

\subsection{Stochastic Discrepancy Oracle}
\label{sec:oracle}

\tool{QRisk} uses a discrepancy oracle to assess whether circuit reduction lowers hardware error relative to the model prediction for the same reduced circuit.
This oracle instantiates the $\mathrm{Drop}$ and $\mathrm{Sufficient}$ predicates used by delta debugging (Section~\ref{sec:ddmin}).
The oracle has three components: a ratio-based design that normalizes observed hardware error by the model-predicted error for the same reduced circuit, a shot-noise threshold calibrated per circuit, and a denominator floor that prevents spurious signals when the predicted error becomes too small.

\noindent \textbf{Ratio-based oracle.} In classical delta debugging, the oracle answers: \emph{does the reduced input still fail?}
In \tool{QRisk}, the oracle answers: \emph{does the reduced circuit still show excess hardware error beyond what the noise model predicts?}

Each oracle call samples three distributions with fresh shots for the reduced circuit $C'$: ideal simulation $P_{\text{ideal}}(C')$, noisy simulation $P_{\text{noisy}}(C')$ using the backend's calibration data, and hardware execution $P_{\text{hw}}(C')$.
We define two distance measures:
\begin{itemize}
\item $\Delta_{\text{model}} = \mathrm{TVD}(P_{\text{ideal}}, P_{\text{noisy}})$: the error predicted by the calibration-based noise model.
\item $\Delta_{\text{mismatch}} = \mathrm{TVD}(P_{\text{noisy}}, P_{\text{hw}})$: the hardware--model mismatch.
\end{itemize}
The oracle computes the discrepancy ratio:
\[
R(C') = \frac{\mathrm{TVD}(P_{\text{ideal}}(C'),\; P_{\text{hw}}(C'))}{\mathrm{TVD}(P_{\text{ideal}}(C'),\; P_{\text{noisy}}(C'))}.
\]
The triangle inequality gives the lower bound
\[
\Delta_{\text{mismatch}} \geq \bigl(R(C')-1\bigr)\Delta_{\text{model}}.
\]
The ratio compares observed hardware error with the model-predicted error for the same reduced circuit.
This normalization helps distinguish changes associated with circuit reduction from excess error beyond the model prediction.
A ratio $R > 1$ indicates greater distance from the ideal distribution on hardware than in the noisy simulation.

\noindent \textbf{Shot-noise threshold.} The ratio $R$ fluctuates due to finite sampling.
We estimate an empirical sampling-noise threshold once on the full input circuit before the DDMin loop, using five pairs of independent noisy simulations:
\[
\tau = \max\bigl(0,\; \overline{R_{\text{null}}} + 2\,\sigma(R_{\text{null}}) - 1.0\bigr),
\]
where $R_{\text{null}} = \mathrm{TVD}(P_{\text{ideal}}, P_{\text{noisy,2}}) / \mathrm{TVD}(P_{\text{ideal}}, P_{\text{noisy,1}})$.
Both distributions come from the same noise model and same circuit, so deviations from 1.0 are purely shot noise.
The threshold remains fixed throughout the run and is used to evaluate ratio drops and elevated retained-circuit ratios.

\noindent \textbf{Denominator floor.} As DDMin removes segments, the denominator $\mathrm{TVD}(\text{ideal}, \text{noisy})$ can shrink.
When it approaches shot-noise levels, small fluctuations cause large ratio swings.
We set a floor: $\mathrm{TVD}_{\min} = 2 \times \overline{\mathrm{TVD}(\text{noisy}_1, \text{noisy}_2)}$.
The floor is estimated from the same full-circuit simulations and remains fixed throughout the run.
If a tested circuit falls below the floor, that call does not trigger either predicate; DDMin proceeds to the next test.

\subsection{Segment-Level Isolation with Delta Debugging}
\label{sec:ddmin}

\begin{figure}[!t]
  \centering
  \includegraphics[width=0.5\textwidth]{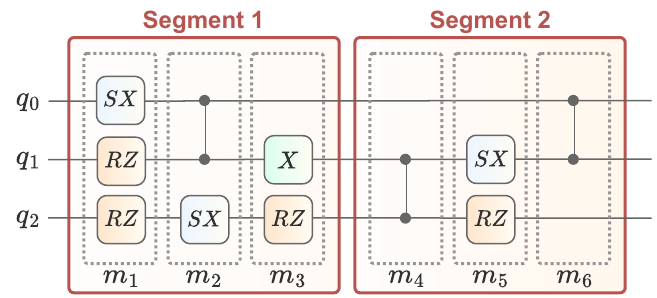}
  \caption{A 3-qubit circuit partitioned into moments and segments.
  Each moment (dashed box) groups gates that execute in parallel.
  Consecutive moments form segments, the atomic units for delta debugging reduction. Each gate is placed in the earliest moment in which all of its qubits are free (ASAP layering), and each segment covers three consecutive moments.}
  \Description{A three-qubit circuit diagram divided into moment columns grouped into segments for DDMin reduction.}
  \label{fig:moments}
\end{figure}

\noindent \textbf{Circuit segmentation.} We partition the circuit by \emph{moments}: groups of gates that can execute in parallel (each gate is assigned to the earliest moment in which all its qubits are free, known as ASAP layering).
Consecutive moments are grouped into fixed-size \emph{segments}, each serving as one atomic unit for DDMin to remove or retain.
This granularity preserves short multi-gate neighborhoods while allowing DDMin to localize hardware--model mismatch.
The segment size is configurable (see Section~\ref{sec:setup}).
Partitioning is a single linear-time pass over the compiled circuit and applies to any gate-model circuit without manual input.
Segments do not overlap because DDMin's complement test requires a partition of the input. A pattern that crosses a segment boundary is still captured when DDMin retains the adjacent segments together. Running \tool{QRisk} at several segment offsets adds overlap without breaking the partition.
Figure~\ref{fig:moments} illustrates this partitioning.

\noindent \textbf{Adapted DDMin loop.} DDMin maintains a candidate set of segments and a partition count $n$.
It divides the candidate set into groups and tests the circuit with each group removed.
If the ratio drops by more than $\tau$, the removed group is implicated in the discrepancy of the tested circuit and becomes the next candidate set.
The same rule applies when removing the complement.
If neither removal test identifies a candidate, Sufficient reuses the complement-removal measurement to check whether $R>1+\tau$.
For candidate set $A$ and group $G$, this test removes $A\setminus G$ from the full input circuit; gates outside $A$ remain.
The search continues until a single segment remains or the partition limit prevents further subdivision.
Both decisions select candidates within the tested circuit context.
At termination, the implementation returns the selected candidates without an additional standalone test.
The capped search identifies candidate segments without guaranteeing classical 1-minimality.
Algorithm~\ref{alg:ddmin} gives the procedure.
\begin{algorithm}[!t]
\caption{DDMin adapted for quantum circuits. $\mathrm{Drop}(X)$ tests whether removing $X$ lowers $R$ by more than $\tau$ relative to the full-circuit baseline. $\mathrm{Sufficient}(X;A)$ reuses the test removing $A\setminus X$ and checks $R>1+\tau$. Both predicates require the denominator floor. The partition limit $n_{\max}$ bounds subdivision.}
\label{alg:ddmin}

\centering
\scalebox{0.8}{%
  \begin{minipage}{1.25\linewidth}
    \begin{algorithmic}[1]
      \REQUIRE Segments $S = \{s_1, \ldots, s_m\}$, predicates $\mathrm{Drop}$, $\mathrm{Sufficient}$
      \ENSURE Candidate abnormal segment set
      \STATE $\textit{cands} \leftarrow S$,\; $n \leftarrow 2$
      \WHILE{$|\textit{cands}| \geq 2$}
        \STATE Split \textit{cands} into $\min(n, |\textit{cands}|)$ non-empty groups $G_1, \ldots, G_k$
        \FOR{each group $G_i$}
          \IF{$\mathrm{Drop}(G_i)$}
            \STATE $\textit{cands} \leftarrow G_i$,\; $n \leftarrow 2$;\; \textbf{restart while}
          \ELSIF{$\mathrm{Drop}(\textit{cands} \setminus G_i)$}
            \STATE $\textit{cands} \leftarrow \textit{cands} \setminus G_i$,\; $n \leftarrow 2$;\; \textbf{restart while}
          \ELSIF{$\mathrm{Sufficient}(G_i;\textit{cands})$ \AND $|G_i| < |\textit{cands}|$}
            \STATE $\textit{cands} \leftarrow G_i$,\; $n \leftarrow 2$;\; \textbf{restart while}
          \ENDIF
        \ENDFOR
        \STATE $n \leftarrow 2n$;\; \textbf{if} $n > n_{\max}$ \textbf{then return} \textit{cands}
      \ENDWHILE
      \RETURN \textit{cands}
    \end{algorithmic}
  \end{minipage}%
}
\end{algorithm}

\subsection{Cross-Window Recurrence Validation}
\label{sec:validation}

A single DDMin run does not establish that a pattern recurs across calibration windows.
\tool{QRisk} validates candidate patterns through independent rediscovery across multiple windows.

For the historical RQ1--RQ2 reports, each returned segment contributes a candidate token sequence containing its native gates, physical qubits, and parameters in recorded order.
For example, the returned segments 4--6 in Table~\ref{tab:candidates-fez} contain 13 gates in total, and segment 5 supplies the four-gate candidate independently rediscovered in eight observations.
Validation uses independent rediscovery in the tested circuits, without separately executing the stored token sequence as a standalone program.
Pattern identity is defined by the backend, physical qubit set, ordered native-gate subsequence, and gate parameters.
A match requires the same gates with matching parameters on the same qubits in the same order.

For each layout under observation, \tool{QRisk} runs the full DDMin isolation procedure independently once per calibration window.
Each run performs its own baseline measurement and threshold calibration, then searches for abnormal patterns from scratch.
A candidate enters the backend-specific pattern database only when DDMin independently isolates a matching fragment in at least one additional non-overlapping window.
We call a pattern \emph{persistent} when it is independently rediscovered in at least two non-overlapping windows.
This criterion follows AlGhmadi et al.'s definition of repetitive patterns in non-stationary systems~\cite{alghmadi2016automated}: a pattern is repetitive when a matching observation can be found in another non-overlapping period.

\section{Evaluation}
\label{sec:evaluation}

This section evaluates \tool{QRisk}'s pattern discovery, recurrence validation, and utility for compilation.
Section~\ref{sec:setup} describes the experimental setup.
Sections~\ref{sec:eval-rq1} and~\ref{sec:eval-rq2} examine pattern isolation and cross-window recurrence, respectively.
Section~\ref{sec:eval-rq3} evaluates pattern-guided circuit rewrites, and Section~\ref{sec:eval-rq4} extends discovery to benchmark workloads.
Section~\ref{sec:eval-cost} reports discovery and compilation costs.

\subsection{Experimental Setup}
\label{sec:setup}

\noindent \textbf{Hardware and circuits.}
We use three IBM Heron r2 backends: \texttt{ibm\_fez}, \texttt{ibm\_marrakesh}, and \texttt{ibm\_kingston}, all 156-qubit Heron r2 processors~\cite{abughanem2025ibm}.
\texttt{ibm\_kingston} also serves as a cross-backend transfer test.
All circuits are compiled with Qiskit~\cite{qiskit2024,qiskit13transpiler} at optimization level~3, the highest preset optimization level, with native gates \{\texttt{rz}, \texttt{sx}, \texttt{x}, \texttt{cz}\}.
We use 3-qubit Grover search~\cite{grover1996fast} with one ancilla (4 physical qubits) as the discovery workload in RQ1--RQ3.
RQ1--RQ2 use one Grover iteration; RQ3 and the Grover reference in RQ4 use three. Table~\ref{tab:setup} gives the corresponding gate and segment counts. Native-gate counts exclude barriers and measurements.
 RQ4 (Section~\ref{sec:eval-rq4}) tests discovery on benchmark workloads from MQT~Bench~\cite{quetschlich2023mqtbench}.
We run \tool{QRisk} on 30 distinct qubit layouts: 10 on \texttt{ibm\_fez}, 10 on \texttt{ibm\_marrakesh}, and 10 on \texttt{ibm\_kingston}.
Each layout is a connected 4-qubit star on the heavy-hex lattice, a degree-3 qubit and its three neighbors, which hosts three data qubits and the ancilla. Seven of the ten positions (2/3/4/16, 24/25/26/37, 50/51/52/58, 64/65/66/77, 100/101/102/116, 126/127/128/137, and 138/150/151/152) are identical on all three backends, enabling same-position comparison. The remaining three per backend cover other regions of each chip.
Patterns from all 30 layouts undergo cross-window validation by running DDMin independently in each observation window; Section~\ref{sec:eval-rq2} reports how many patterns pass validation (fragments recurring across independent DDMin runs).

\noindent \textbf{Experimental design.}
\emph{Discovery and validation (RQ1--RQ2).}
Each layout undergoes one DDMin run per observation window. We partition each circuit into 3-moment segments, and DDMin searches over at most 16 partitions (Section~\ref{sec:ddmin}).
Each oracle query compares three output distributions, each sampled with 8{,}192 fresh shots: ideal distribution from noiseless Aer simulation, noisy distribution from Aer simulation using the backend's most recent calibration~\cite{qiskitaer2025noisemodel}, and the distribution measured on the physical device.
A candidate fragment is retained as a persistent pattern only when DDMin independently rediscovers it in at least one additional non-overlapping window (Section~\ref{sec:validation}).
Gate parameters are matched with a tolerance of $10^{-5}$ radians, following the pattern database implementation; all parameters originate from Qiskit’s compiler.

\emph{Controlled intervention (RQ3).}
For each of the two selected patterns, we start from the Qiskit O3-compiled circuit and construct 40 semantically equivalent variants: 10 variants at each of four pattern counts (0--3).
The variants preserve gate composition and have similar depth.
Commuting gate swaps vary pattern count to test whether breaking these sequences reduces hardware--model mismatch.
The unmodified O3 output of the three-iteration Grover circuit contains each pattern exactly three times, once per iteration, providing the reference gate ordering for the three-occurrence level. Variants are produced by applying random commuting swaps with varying swap counts and seeds. Each variant is assigned to a level by the same exact-match scanner used for the pattern database, variants whose moment structure is unchanged are discarded, and variants are sampled so that the number of swaps is balanced across levels.
We additionally run the \texttt{ibm\_fez} variants on \texttt{ibm\_kingston} as a cross-backend transfer test (Section~\ref{sec:cross-backend}).

\noindent \textbf{Metrics and statistical tests.}
The primary metric is hardware--model mismatch, $\Delta_{\text{mismatch}}=\mathrm{TVD}(\text{noisy}, \text{real})$ (Section~\ref{sec:oracle}).
We also report the discrepancy ratio $R$ and pattern occurrence counts.
The RQ1--RQ3 circuits use four physical qubits and measure three data qubits, yielding eight output outcomes.
The denominator floor is estimated from independent noisy simulations of the same circuit, as defined in Section~\ref{sec:oracle}.
Cross-window recurrence is measured as the fraction of windows in which a segment is flagged.
In RQ3, Spearman's $\rho$ tests the association between pattern count and mismatch.
The one-sided Mann--Whitney $U$ test compares the ten zero-occurrence variants with the ten three-occurrence variants, testing whether mismatch is lower in the zero-occurrence group.
We use $p < 0.01$ as the significance threshold.
Table~\ref{tab:setup} summarizes the experimental settings.

\begin{table}[!t]
\centering
\caption{Experimental setup summary. The hardware, compilation, and oracle settings apply to RQ1--RQ3; the RQ3 and RQ4 blocks specify workload and protocol differences.}
\label{tab:setup}

\footnotesize
\setlength{\tabcolsep}{4pt}
\renewcommand{\arraystretch}{1.00}

\begin{tabularx}{\columnwidth}{
    @{}c|
    >{\raggedright\arraybackslash}p{0.31\columnwidth}|
    >{\raggedright\arraybackslash}X
    @{}
}
\toprule
& \textbf{Parameter} & \textbf{Value} \\
\midrule

\multirow{6}{*}{%
  \rotatebox[origin=c]{90}{%
    \scriptsize
    \shortstack{
      \textit{Hardware and}\\
      \textit{compilation}
    }
  }
}
& Backends
& \texttt{ibm\_fez}, \texttt{ibm\_marrakesh}, \texttt{ibm\_kingston} \\

& Backend architecture
& IBM Heron r2 (156 qubits) \\

& Qiskit version
& 1.3.1 \\

& Aer version
& 0.15.1 \\

& Optimization level
& 3 (Qiskit O3) \\

& Native gate set
& \{rz, sx, x, cz\} \\

\midrule

\multirow{4}{*}{%
  \rotatebox[origin=c]{90}{%
    \scriptsize
    \shortstack{
      \textit{Study}\\
      \textit{design}
    }
  }
}
& Total layouts
& 30 (10 per backend) \\

& Long-term observation
& 8 months (Nov 2025--Jun 2026) \\

& Post-update observation
& May 6--June 25, 2026 (about 7 weeks) \\

& Workload
& One-iteration Grover + ancilla (RQ1--RQ2): 151/158 native gates, 36/37 segments; three iterations in RQ3 \\

\midrule

\multirow{7}{*}{%
  \rotatebox[origin=c]{90}{%
    \scriptsize
    \shortstack{
      \textit{Oracle and DDMin}\\
      \textit{parameters}
    }
  }
}
& Shots per distribution
& 8{,}192 \\

& Segment size
& 3 moments \\

& Max partitions
& 16 \\

& Sigma multiplier
& 2 \\

& Denominator floor
& $2 \times \overline{\mathrm{TVD}(\text{noisy}_1,\text{noisy}_2)}$ \\

& Recurrence threshold
& $\geq 2$ windows \\

& Parameter precision
& $10^{-5}$ rad \\

\midrule

\multirow{6}{*}{%
  \rotatebox[origin=c]{90}{%
    \tiny
    \shortstack{
      \textit{RQ3}\\
      \textit{controlled}\\
      \textit{intervention}
    }
  }
}
& Execution date
& June 2, 2026 \\

& Variants per layout
& 40 (10 per occurrence level) \\

& Pattern occurrence levels
& 0, 1, 2, 3 \\

& Variant construction
& Commuting gate swaps \\

& Native gates / segments
& 453/107 (Fez), 474/111 (Marrakesh) \\

& Instruction convention
& 456/477 includes three measurement instructions \\

\midrule

\multirow{6}{*}{%
  \rotatebox[origin=c]{90}{%
    \scriptsize
    \shortstack{
      \textit{RQ4}\\
      \textit{cross-benchmark}\\
      \textit{study}
    }%
  }
}
& Backend / date
& \texttt{ibm\_fez}, one calibration window (2026-09-25) \\

& Layouts
& 46/47/48/57, 97/106/107/108 \\

& Workloads
& Three-iteration Grover (431/436 native gates; 105/106 segments) + 13 MQT Bench 2.3.0 circuits \\

& Compiler
& Qiskit 2.4.1, O3, \texttt{seed\_transpiler}=7 \\

& Oracle-call cap
& 30 per run; segment 0 not selectable \\

& Hardware cost
& 192 executions, 12.5 QPU-minutes \\

\bottomrule
\end{tabularx}

\end{table}

\subsection{RQ1: Can \tool{QRisk} isolate compact abnormal patterns from compiled quantum circuits across diverse backends and layouts?}
\label{sec:eval-rq1}

We assess whether DDMin can isolate compact candidate patterns across backends and qubit layouts.
\tool{QRisk} discovers 92 distinct candidate patterns from 30 layouts covering 117 unique qubits across three backends.
Table~\ref{tab:inventory} shows representative rediscovered candidates and single-window candidates from both long-term and post-update observations.
Each pattern spans 3 to 6 gates on 2 to 4 coupled qubits, and DDMin converges in 5 to 8 steps from circuits of 151 to 158 operations.
Section~\ref{sec:eval-rq2} analyzes which of these patterns recur across windows.

\begin{table*}[!t]
  \centering
  \small
  \caption{Representative candidate patterns across 30 tested layouts on 3 backends. Among 92 distinct patterns discovered, 25 pass cross-window validation (recurrence $\geq$ 2), while 67 produce only single-window candidates. Recurrence counts require exact match of backend, physical qubits, gate order, and gate parameters; patterns shown as gate-type projection for readability. First/last seen dates show when each pattern was first/last independently rediscovered. Mean $R$ is the converged discrepancy ratio of the tested reduced circuit recorded by the run ($R$, Section~\ref{sec:oracle}), averaged over the pattern's rediscovery windows.}
  \label{tab:inventory}

  \resizebox{\textwidth}{!}{%
    \begin{tabular}{lcccccll}
      \toprule
      Detected pattern (gate-type projection) & First seen & Last seen & Rec. & Mean $R$ & Obs. period & Backend & Layout \\
      \midrule
      \multicolumn{8}{l}{\emph{Long-term observation (Nov 2025--Jun 2026, monthly to biweekly windows)}} \\
      rz $\to$ cz $\to$ sx                    & Nov 18 & Apr 13 & 10 & 1.15 & 8 months & marrakesh & 6/7/8/17 \\
      sx $\to$ sx $\to$ rz $\to$ cz           & Nov 15 & Mar 12 & 8 & 1.19 & 8 months & fez       & 97/106/107/108 \\
      rz $\to$ rz $\to$ sx $\to$ cz           & Nov 15 & Mar 4  & 7 & 1.21 & 8 months & fez       & 97/106/107/108 \\
      rz $\to$ sx $\to$ cz                    & Nov 18 & Apr 9  & 6 & 1.17 & 8 months & marrakesh & 6/7/8/17 \\
      \multicolumn{8}{l}{\emph{... and 8 additional patterns (12 from long-term observation total)}} \\
      \midrule
      \multicolumn{8}{l}{\emph{Post-update observation (May--Jun 2026, weekly windows)}} \\
      rz $\to$ cz $\to$ sx                    & May 20 & Jun 24 & 4 & 1.54 & May 6--Jun 24 & fez       & 46/47/48/57 \\
      cz $\to$ sx $\to$ rz                    & May 7  & Jun 17 & 4 & 1.63 & May 7--Jun 25  & kingston  & 138/150/151/152 \\
      cz $\to$ rz $\to$ sx $\to$ sx           & Jun 3  & Jun 17 & 3 & 1.37 & May 6--Jun 24  & marrakesh & 118/128/129/130 \\
      sx $\to$ sx $\to$ rz $\to$ rz $\to$ sx $\to$ rz
                                               & May 27 & Jun 10 & 3 & 1.61 & May 7--Jun 24  & marrakesh & 6/7/8/17 \\
      \multicolumn{8}{l}{\emph{... and 9 additional patterns (13 from post-update observation total)}} \\
      \midrule
      \multicolumn{8}{l}{\emph{Examples of candidates not rediscovered (67 with single occurrence)}} \\
      sx $\to$ rz $\to$ sx                    & Nov 29 & -- & 1 & 1.12 & 8 months & marrakesh & 6/7/8/17 \\
      rz $\to$ sx $\to$ cz                    & Dec 10 & -- & 1 & 1.35 & 8 months & fez       & 97/106/107/108 \\
      x $\to$ sx $\to$ rz $\to$ rz $\to$ sx   & May 13 & -- & 1 & 4.19 & May 7--Jun 24 & marrakesh & 2/3/4/16 \\
      rz $\to$ sx $\to$ cz                    & May 20 & -- & 1 & 1.81 & May 7--Jun 24 & kingston  & 126/127/128/137 \\
      \bottomrule
    \end{tabular}%
  }

\end{table*}

\indent \textbf{DDMin reduction example.}
On \texttt{ibm\_fez} qubits 97/106/107/108, the O3-compiled Grover circuit has 151 operations over 107 moments (36 segments) and a baseline ratio $R = 1.3$, which exceeds $1+\tau$ (Section~\ref{sec:oracle}).
DDMin returns three adjacent candidate segments (4--6; Table~\ref{tab:candidates-fez}).
The DDMin reduction process tests candidate subsets at each step, measuring the ratio and deciding whether to keep or subdivide based on whether the ratio remains elevated.
Table~\ref{tab:ddmin-trace-fez} shows selected tests recorded for this reduction.
No single CZ qubit pair is disproportionately represented in the abnormal segments (representation ratios 0.67x--1.26x relative to the full circuit's CZ distribution), so no single pair is substantially overrepresented.
The \texttt{ibm\_marrakesh} layout (qubits 6/7/8/17, 158 operations, 37 segments, $R = 1.38$) shows the same pattern: DDMin isolates contiguous abnormal segments that recur across windows (Table~\ref{tab:inventory}).

\begingroup

\captionsetup{
  width=0.76\columnwidth,
  position=top,
  font=tablefont,
  labelfont={tablelabel,bf},
  labelformat=spacedtable,
  labelsep=spacedperiod,
  justification=raggedright,
  singlelinecheck=false,
  skip=4pt
}

\newsavebox{\circuitfour}
\sbox{\circuitfour}{%
\fontsize{6}{7}\selectfont
\begin{quantikz}[column sep=0.25cm,row sep=0.05cm]
\lstick{q97}  & \gate{RZ(-2.75)} & \qw & \qw & \qw & \qw \\
\lstick{q107} & \qw & \gate{RZ(2.77)} & \gate{SX} & \ctrl{1} & \qw \\
\lstick{q108} & \qw & \qw & \qw & \ctrl{-1} & \qw
\end{quantikz}}

\newsavebox{\circuitfive}
\sbox{\circuitfive}{%
\fontsize{6}{7}\selectfont
\begin{quantikz}[column sep=0.25cm,row sep=0.05cm]
\lstick{q107} & \gate{SX} & \qw & \gate{RZ(0.392699)} & \ctrl{1} & \qw \\
\lstick{q108} & \qw & \gate{SX} & \qw & \ctrl{-1} & \qw
\end{quantikz}}

\newsavebox{\circuitsix}
\sbox{\circuitsix}{%
\fontsize{6}{7}\selectfont
\begin{quantikz}[column sep=0.25cm,row sep=0.05cm]
\lstick{q97}  & \qw & \qw & \ctrl{1} & \qw & \qw & \qw \\
\lstick{q107} & \gate{SX} & \qw & \ctrl{-1} & \qw & \gate{SX} & \qw \\
\lstick{q108} & \qw & \gate{SX} & \qw & \gate{RZ(0.76)} & \qw & \qw
\end{quantikz}}

\begin{table}[!ht]
\centering
\caption{Recorded DDMin tests on \texttt{ibm\_fez} ($\tau=0.06$). Narrow selects the removed group when its removal reduces $R$ by more than $\tau$ relative to the full-circuit baseline. Sufficient selects a retained group when the complement-removal test has $R>1+\tau$. Typical discovery runs use about 12--14 oracle calls (Table~\ref{tab:cost}).}
\label{tab:ddmin-trace-fez}
\small
\renewcommand{\arraystretch}{1.35}
\setlength{\tabcolsep}{10pt}
\begin{tabular}{@{}clrl@{}}
\toprule
Record & Tested circuit & $R$ & Action \\
\midrule
1 & Full circuit & 1.30 & Start \\
2 & Remove 0--17 & 1.05 & Narrow \\
3 & Remove 0--8 & 1.22 & Narrow \\
4 & Remove 0--6 & 1.19 & Narrow \\
5 & Retain 4--6 & 1.15 & Sufficient \\
\bottomrule
\end{tabular}
\end{table}

\begin{table}[!ht]
\centering
\caption{Candidate segments returned on \texttt{ibm\_fez}. The input circuit has 151 operations and 36 segments.}
\label{tab:candidates-fez}
\small
\renewcommand{\arraystretch}{1.15}
\begin{tabularx}{0.76\columnwidth}{@{}c>{\centering\arraybackslash}X@{}}
\toprule
Segment & Gate sequence \\
\midrule
4 & \usebox{\circuitfour} \\
\midrule
5 & \usebox{\circuitfive} \\
\midrule
6 & \usebox{\circuitsix} \\
\bottomrule
\end{tabularx}
\end{table}

\endgroup

Of the 205 DDMin runs that return candidates, 176 terminate with one segment and 17 return multi-segment sets through a Sufficient decision.
These 193 runs account for 94.1\% of the candidate-returning runs; the remaining 12 return multi-segment sets.

\finding{\tool{QRisk} reduces 151--158 operation circuits to 3--6 gate candidate patterns in 5--8 DDMin steps across 30 layouts on three IBM Heron r2 backends. Across 205 candidate-returning runs, 193 (94.1\%) end with one segment or a Sufficient decision.}

\subsection{RQ2: How effectively does cross-window validation identify persistent hardware patterns?}
\label{sec:eval-rq2}

RQ2 evaluates cross-window validation by measuring how many candidates satisfy the persistence criterion: independent rediscovery in at least two non-overlapping calibration windows.
For each of the 30 layouts, we apply the pattern identity and validation procedure defined in Section~\ref{sec:validation}.

\indent \textbf{Patterns retained by validation.}
Among the 92 distinct patterns discovered, 25 (27\%) pass validation through independent rediscovery in multiple windows.
The remaining 67 candidates (73\%) are observed in one window and are not rediscovered within the observation period.
Table~\ref{tab:inventory} shows examples from both groups.
Among the 258 recorded DDMin runs, 205 return candidate segments; the other 53 reports contain no returned candidate.

\indent \textbf{Long-term observation (monthly to biweekly windows).}
12 patterns discovered from two layouts observed from November 2025 through June 2026 (fez 97/106/107/108 and marrakesh 6/7/8/17, shown in Table~\ref{tab:inventory}) demonstrate recurrence across multiple non-overlapping windows and calibration cycles during the 8-month observation period.

\indent \textbf{Post-update observation (weekly windows).}
After IBM's hardware update at the end of April 2026, \tool{QRisk} discovered and validated 13 new patterns from eight layouts among the 30 observed during May--June, including layouts already under long-term observation.
Each pattern was independently rediscovered in multiple weekly windows.
The post-update reports span May 6--June 25, 2026; Table~\ref{tab:inventory} gives the observation dates for each example.

\indent \textbf{Pattern lifetime.}
For the two layouts observed over 8 months, the highest-recurrence patterns were rediscovered from November 2025 through March--April 2026 but not in the subsequent May--June observations.
DDMin isolated different segment sets in the later observations, indicating a change in the flagged gate sequences.
These observations motivate periodic re-validation of stored patterns.

\finding{Cross-window validation retains 25 persistent patterns (27\% of the 92 candidates), providing repeated evidence for their inclusion in the backend-specific pattern database.}

\subsection{RQ3: How can these patterns help quantum compilers?}
\label{sec:eval-rq3}

\indent \textbf{Pattern-guided rewrite.}
\label{sec:application}
A commuting-gate swap (Section~\ref{sec:qc-compilation}) reorders an adjacent commuting pair such as \texttt{rz} and \texttt{cz}.
When a matched pattern contains an adjacent commuting pair, swapping that pair changes the matched gate order while preserving circuit semantics.
The rewrite is implemented as a Qiskit \texttt{TransformationPass} that scans a compiled circuit for database matches and applies the swaps at compile time, with no additional hardware executions.

\begin{figure*}[!t]
  \centering
  \begin{subfigure}[t]{0.32\textwidth}
    \centering
    \includegraphics[width=\textwidth]{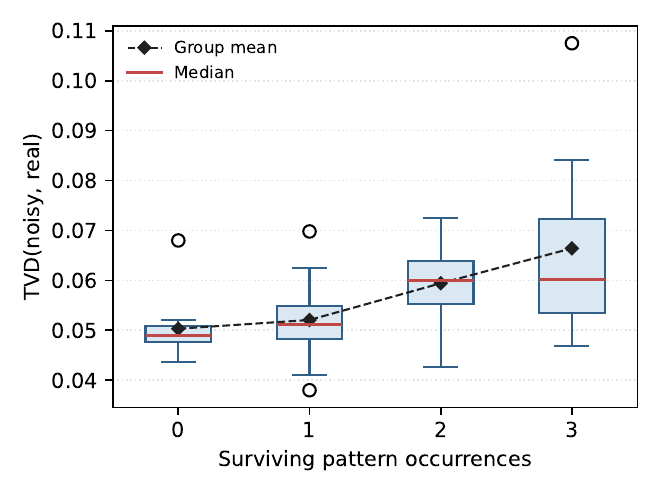}
    \caption{\texttt{ibm\_fez}: mean $-$24\%.}
    \Description{Scatter and trend plot for ibm_fez showing lower TVD as the abnormal pattern is systematically broken, reducing occurrences from three to zero.}
    \label{fig:scaling-fez}
  \end{subfigure}
  \hfill
  \begin{subfigure}[t]{0.32\textwidth}
    \centering
    \includegraphics[width=\textwidth]{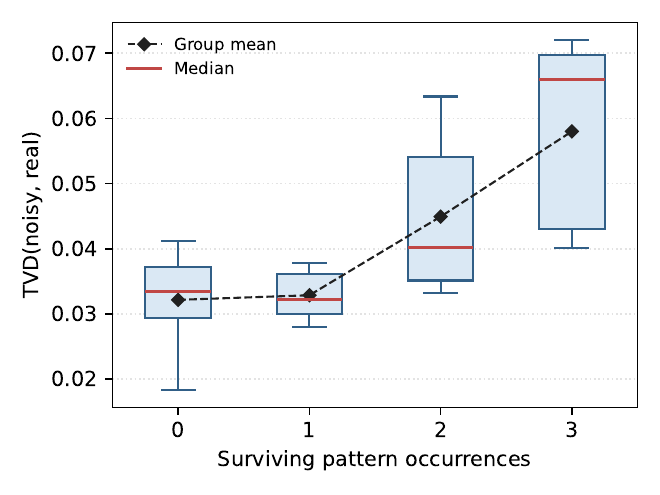}
    \caption{\texttt{ibm\_marrakesh}: mean $-$45\%.}
    \Description{Scatter and trend plot for ibm_marrakesh showing lower TVD as the abnormal pattern is systematically broken, reducing occurrences from three to zero.}
    \label{fig:scaling-marrakesh}
  \end{subfigure}
  \hfill
  \begin{subfigure}[t]{0.32\textwidth}
    \centering
    \includegraphics[width=\textwidth]{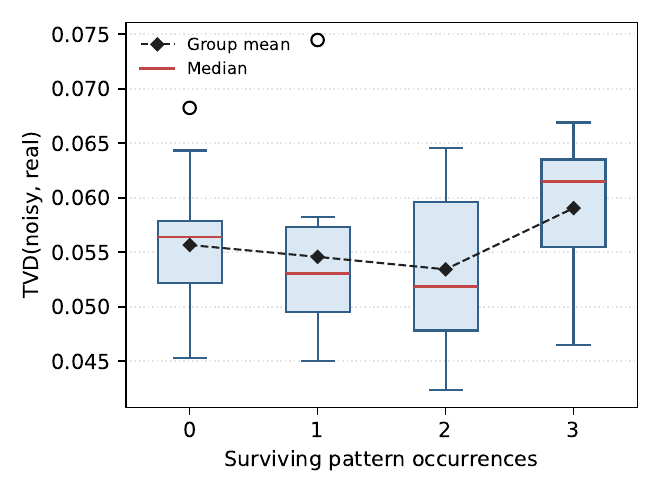}
    \caption{\texttt{ibm\_kingston}: no monotonic trend.}
    \Description{Scatter and trend plot for ibm_kingston showing no consistent relationship between abnormal-pattern count and TVD.}
    \label{fig:scaling-kingston}
  \end{subfigure}
  \caption{Controlled intervention: $\mathrm{TVD}(\text{noisy}, \text{real})$ vs.\ surviving pattern count (10 circuits per group). Black diamonds and dashed lines mark group means; red boxplot lines mark medians. (a) and (b) show mean hardware--model mismatch reductions of 24\% and 45\%, respectively, from three to zero pattern occurrences after commuting-gate reordering. (c) runs the same experiment from \texttt{fez} on a different backend \texttt{kingston}, showing no statistically significant correlation between pattern count and hardware--model mismatch on \texttt{kingston}.}
  \label{fig:scaling}
\end{figure*}

\indent \textbf{Effect of pattern removal.}
We test whether the discovered patterns predict hardware--model mismatch by systematically varying their occurrence count in compiled circuits.
Using the two highest-recurrence patterns from long-term observation in Table~\ref{tab:inventory} (\texttt{ibm\_fez} 97/106/107/108, pattern: sx $\to$ sx $\to$ rz $\to$ cz, recurrence 8; \texttt{ibm\_marrakesh} 6/7/8/17, pattern: rz $\to$ cz $\to$ sx, recurrence 10), we generate 40 variants per backend (10 per occurrence level, 0 to 3 surviving pattern occurrences) that preserve gate composition and have similar depth.
All variants for a given backend share identical instruction counts (456 instructions for \texttt{ibm\_fez}, 477 for \texttt{ibm\_marrakesh}, including three measurements) and nearly identical circuit depth (316--322 for fez and 326--333 for marrakesh).
The variants differ only in gate sequence through commuting swaps that reorder independent operations without introducing or removing gates.
Zero-occurrence variants serve as the comparison group where the pattern has been broken by commuting gate swaps.
Execution order is shuffled to interleave the occurrence levels.
In the unmodified O3 output, the three occurrences sit at 8\%, 41\%, and 74\% of the gate sequence, one per Grover iteration.
The variants therefore break occurrences at early, middle, and late positions in the circuit.
Breaking patterns through commuting gate swaps reduces mean hardware--model mismatch on \texttt{ibm\_fez}: $0.066 \to 0.059 \to 0.052 \to 0.050$ in $\mathrm{TVD}(\text{noisy}, \text{real})$, a 24\% reduction from three to zero occurrences (Figure~\ref{fig:scaling-fez}).
On \texttt{ibm\_marrakesh} the same relationship holds: $0.058 \to 0.045 \to 0.033 \to 0.032$, a 45\% reduction (Figure~\ref{fig:scaling-marrakesh}).
The monotone trend in the group means supports the hypothesis that the pattern contributes to hardware--model mismatch.
A second outcome measure, $\mathrm{TVD}(\text{ideal}, \text{real})$, yields the same conclusion. Pattern count correlates with end-to-end error on \texttt{fez} ($\rho = 0.437$, $p = 0.005$) and \texttt{marrakesh} ($\rho = 0.595$, $p < 0.0001$), but not on \texttt{kingston}. Eliminating all pattern occurrences also lowers end-to-end $\mathrm{TVD}(\text{ideal}, \text{real})$ by 12\% on \texttt{fez} and 19\% on \texttt{marrakesh}.

Across variants with the same gate composition, physical qubits, and per-gate noise-model score, and similar depth, commuting-gate reordering reduces mean hardware--model mismatch on \texttt{ibm\_fez} and \texttt{ibm\_marrakesh} (Table~\ref{tab:stats}).
These results support using execution-derived gate-ordering information to complement the calibration-based Aer model.

\begin{table*}[!t]
  \centering
  \small
  \caption{Statistical tests for the controlled intervention ($N{=}40$ per backend). Mann--Whitney compares 0 vs. 3 occurrences, testing for lower mismatch with 0 occurrences. Significant results ($p < 0.01$) are bold.}
  \label{tab:stats}
   \scalebox{0.8}{
  \begin{tabular}{l cccc cc}
    \toprule
    & \multicolumn{4}{c}{Pattern count vs.\ TVD(noisy,real)} & \multicolumn{2}{c}{Pattern count vs.\ TVD(ideal,noisy)} \\
    \cmidrule(lr){2-5} \cmidrule(lr){6-7}
    Backend & Spearman $\rho$ & $p$ & M-W $U$ & $p$ (1-sided) & Spearman $\rho$ & $p$ \\
    \midrule
    \texttt{ibm\_fez} & \textbf{0.515} & \textbf{0.0007} & \textbf{13.0} & \textbf{0.003} & 0.158 & 0.33 \\
    \texttt{ibm\_marrakesh} & \textbf{0.711} & $\mathbf{<0.0001}$ & \textbf{2.0} & \textbf{0.0002} & $-0.066$ & 0.69 \\
    \texttt{ibm\_kingston} (control) & 0.187 & 0.248 & 34.0 & 0.121 & 0.059 & 0.72 \\
    \bottomrule
  \end{tabular}
  }
\end{table*}

\finding{Breaking abnormal pattern occurrences from 3 to 0 reduces mean hardware--model mismatch by 24\% on \texttt{fez}
(Spearman $\rho = 0.515$, $p = 0.0007$) and by 45\% on \texttt{marrakesh}  ($\rho = 0.711$, $p < 0.0001$). }

\subsection{RQ4: Which benchmark workloads can \tool{QRisk} use for discovery, and are the fragments it isolates specific to the benchmark?}
\label{sec:eval-rq4}

RQ1--RQ3 use Grover search as the discovery workload.
RQ4 protects segment 0, which contains state preparation, to focus discovery on the subsequent computation. Its 30-call cap bounds hardware executions, whereas the 16-partition cap bounds search subdivision.
RQ4 tests whether discovery extends to other workloads and whether their isolated fragments share the gate sequences found in Grover.

\indent \textbf{Workload screening.}
We screen input circuits using three criteria:
(i)~enough compiled depth to form many segments ($\geq 8$);
(ii)~an ideal output far from uniform, $\mathrm{TVD}(\text{ideal},\text{uniform}) \geq 0.15$, so that hardware and model errors are distinguishable under TVD; and
(iii)~model-predicted error above measurement noise, $\mathrm{TVD}(\text{ideal},\text{noisy}) \geq 1.5\times$ the shot-noise floor.
We compiled Grover-3 and 13 four-qubit MQT~Bench~2.3.0 circuits~\cite{quetschlich2023mqtbench} onto two \texttt{ibm\_fez} layouts and checked each property in simulation using the live calibration (Table~\ref{tab:preflight}); screening uses no hardware.
QFT and graph states produce near-uniform outputs (0.02 from uniform) and fail criterion (ii).
QFT-entangled and inexact QPE have model error at the shot-noise floor, and Deutsch--Jozsa and Bernstein--Vazirani compile to at most 6 segments.
Amplitude estimation (AE), quantum walk, random circuits, W-state, exact QPE, and MQT's Grover pass on both layouts.
Grover is thus a sufficient discovery workload but not a unique one.
We run \tool{QRisk} twice per layout on Grover-3, AE, quantum walk, and random circuits, the qualifying workloads with the most distinct gate structure.
Grover-3 on layout 46/47/48/57 falls just below criterion (iii) (ratio 1.40) but is run as the reference workload.

\begin{table}[!t]
  \centering
  \caption{Workload screening on \texttt{ibm\_fez} (simulation only, live calibration). Values are given as layout 46/47/48/57 / layout 97/106/107/108. ``Model/floor'' is $\mathrm{TVD}(\text{ideal},\text{noisy})$ divided by $2\,\mathrm{TVD}(\text{noisy}_1,\text{noisy}_2)$ from two independent 8{,}192-shot simulations of that workload. This workload-specific divisor ranges from 0.0037 to 0.0681; criterion (iii) requires $\geq 1.5$.}
  \label{tab:preflight}
  \scalebox{0.78}{%
  \begin{tabular}{lccccc}
    \toprule
    Workload & Segments & TVD(ideal, unif.) & Floor & Model/floor & Usable \\
    \midrule
    Grover-3 (this paper) & 105/106 & 0.21 & 0.0364/0.0205 & 1.40/2.11 & --/\checkmark \\
    Amplitude estimation & 22/24 & 0.66 & 0.0266/0.0227 & 2.74/2.47 & \checkmark/\checkmark \\
    Quantum walk & 231/237 & 0.69 & 0.0603/0.0273 & 5.31/11.51 & \checkmark/\checkmark \\
    Random circuit & 98/98 & 0.79 & 0.0349/0.0354 & 5.93/5.30 & \checkmark/\checkmark \\
    W-state & 10/9 & 0.75 & 0.0374/0.0193 & 1.81/2.21 & \checkmark/\checkmark \\
    QPE (exact) & 11/12 & 0.88 & 0.0107/0.0093 & 6.57/5.52 & \checkmark/\checkmark \\
    Grover (MQT) & 118/120 & 0.88 & 0.0449/0.0442 & 8.39/7.76 & \checkmark/\checkmark \\
    GHZ & 8/7 & 0.88 & 0.0149/0.0396 & 5.23/1.00 & \checkmark/-- \\
    QPE (inexact) & 22/22 & 0.57 & 0.0503/0.0422 & 1.13/1.01 & --/-- \\
    QFT (entangled) & 27/25 & 0.32 & 0.0435/0.0527 & 1.02/0.73 & --/-- \\
    QFT & 24/23 & 0.02 & 0.0681/0.0449 & 0.48/0.55 & --/-- \\
    Graph state & 6/5 & 0.02 & 0.0562/0.0566 & 0.52/0.46 & --/-- \\
    Deutsch--Jozsa & 5/6 & 0.88 & 0.0107/0.0071 & 4.38/5.86 & --/-- \\
    Bernstein--Vazirani & 4/2 & 0.88 & 0.0046/0.0037 & 8.78/5.51 & --/-- \\
    \bottomrule
  \end{tabular}}
\end{table}

\begin{table*}[!t]
  \centering
  \caption{\tool{QRisk} runs on \texttt{ibm\_fez} across workloads (one calibration window, two runs per workload and layout). Margin is $R-(1+\tau)$ at baseline; ``Seg.'' is the index of the returned segment (0-based; segment 0 is not selectable). Fragments are the largest qubit-connected components of the returned segment, with physical qubits. Gates are linked when they share a qubit; unrelated single-qubit gates are omitted, and order within the component is retained.}
  \label{tab:xalg-runs}
  \resizebox{\textwidth}{!}{%
  \begin{tabular}{llcccccc l}
    \toprule
    Layout & Workload & Run & Baseline $R$ & $\tau$ & Margin & Oracle calls & Seg. & Isolated fragment \\
    \midrule
    46 & Grover-3 & 1 & 1.87 & 0.080 & +0.79 & 15 & 1 & \texttt{\small rz(46) $\to$ sx(47) $\to$ rz(47) $\to$ cz(46,47)} \\
    46 & Grover-3 & 2 & 2.67 & 0.332 & +1.34 & 12 & 1 & \texttt{\small rz(46) $\to$ sx(47) $\to$ rz(47) $\to$ cz(46,47)} \\
    46 & AE & 1 & 2.74 & 0.157 & +1.58 & 11 & 7 & \texttt{\small sx(57) $\to$ rz(57) $\to$ sx(57)} \\
    46 & AE & 2 & 2.06 & 0.166 & +0.89 & 13 & 9 & \texttt{\small cz(47,48) $\to$ x(47) $\to$ rz(47) $\to$ sx(48) $\to$ rz(48)} \\
    46 & Quantum walk & 1 & 1.42 & 0.064 & +0.35 & 15 & 1 & \texttt{\small cz(47,57) $\to$ sx(47) $\to$ rz(47)} \\
    46 & Quantum walk & 2 & 1.47 & 0.038 & +0.43 & 15 & 1 & \texttt{\small cz(47,57) $\to$ sx(47) $\to$ rz(47)} \\
    46 & Random & 1 & 1.82 & 0.062 & +0.76 & 15 & 1 & \texttt{\small cz(46,47) $\to$ sx(46) $\to$ sx(47) $\to$ cz(46,47)} \\
    46 & Random & 2 & 1.92 & 0.054 & +0.87 & 15 & 1 & \texttt{\small cz(46,47) $\to$ sx(46) $\to$ sx(47) $\to$ cz(46,47)} \\
    \midrule
    97 & Grover-3 & 1 & 1.78 & 0.402 & +0.38 & 14 & 52 & \texttt{\small sx(107) $\to$ cz(97,107) $\to$ sx(107)} \\
    97 & Grover-3 & 2 & 1.93 & 0.480 & +0.45 & 14 & 35 & \texttt{\small sx(106) $\to$ sx(107) $\to$ cz(106,107) $\to$ sx(106) $\to$ sx(107)} \\
    97 & AE & 1 & 1.62 & 0.153 & +0.46 & 12 & 16 & \texttt{\small sx(107) $\to$ rz(107) $\to$ cz(97,107)} \\
    97 & AE & 2 & 1.77 & 0.216 & +0.55 & 12 & 7 & \texttt{\small rz(107) $\to$ cz(106,107) $\to$ x(106) $\to$ sx(107)} \\
    97 & Quantum walk & 1 & 1.39 & 0.030 & +0.36 & 13 & 16 & \texttt{\small cz(106,107) $\to$ sx(107) $\to$ rz(107)} \\
    97 & Quantum walk & 2 & 1.43 & 0.029 & +0.40 & 14 & 16 & \texttt{\small cz(106,107) $\to$ sx(107) $\to$ rz(107)} \\
    97 & Random & 1 & 0.68 & 0.000 & $-$0.32 & 1 & -- & \emph{hardware below model ($R<1$); DDMin not run} \\
    97 & Random & 2 & 0.78 & 0.102 & $-$0.32 & 1 & -- & \emph{hardware below model ($R<1$); DDMin not run} \\
    \bottomrule
  \end{tabular}}
\end{table*}

\indent \textbf{Isolation on hardware.}
Table~\ref{tab:xalg-runs} reports all 16 runs.
In 14 of 16 runs the baseline ratio exceeds $1+\tau$ (margins $+0.35$ to $+1.58$), and DDMin converges to a single segment in 11--15 oracle calls, using 192 hardware executions (12.5 QPU-minutes) in total.
The tested circuits span 22--237 segments.
The connected components have 3--5 gates on 1--2 qubits.

\indent \textbf{Fragment overlap across workloads.}
At the gate-type level the fragments share the native-gate vocabulary, and 13 of 14 involve the layout's center qubit (47 or 107).
This resemblance is expected by construction.
Each layout is a star in which every \texttt{cz} involves the center qubit, and across the eight compiled circuits 92--100\% of all segments touch that qubit and 59--73\% contain all three of \texttt{sx}, \texttt{rz}, and \texttt{cz}.
To assess whether shared couplings exceed what the circuit structure alone would predict, we compare the returned fragments with a null model that selects a segment uniformly at random from each circuit.
Under a looser criterion, sharing an interacting \texttt{cz} pair with a Grover fragment, 6/10 fragments match against 3.51 expected (one-sided Poisson-binomial $p = 0.089$).
The returned fragments share the layout\textquotesingle s gate vocabulary and couplings, while their ordered gate sequences differ across the tested workloads (Table~\ref{tab:xalg-runs}).
Within a layout, repeated runs on the same workload often return the same fragment (Table~\ref{tab:xalg-runs}).

\indent \textbf{Sensitivity to the distance metric.}
To assess whether the observed hardware--model mismatches depend on the distance metric, we recompute the ratios for all 192 hardware executions using the stored raw counts, with Hellinger distance in place of TVD.
The Hellinger ratio correlates strongly with the TVD ratio (Spearman $\rho = 0.874$, $p < 10^{-60}$) and falls on the same side of 1.0 in 99\% of calls, showing close agreement in whether hardware error exceeds the model prediction.

\finding{Six of thirteen MQT~Bench workloads meet all three screening criteria on both layouts.
In the single-window hardware study, \tool{QRisk} isolates 3--5-gate fragments from AE, quantum walk, random circuits, and Grover in 11--15 oracle calls.
The returned fragments share the layout\textquotesingle s native-gate vocabulary and couplings but differ in gate order across workloads.}

\subsection{Discovery and Compilation Costs}
\label{sec:eval-cost}

Table~\ref{tab:cost} summarizes the cost of offline discovery and online use.
Typical RQ1 discovery runs use about 12--14 hardware executions (oracle calls), or 98K--115K shots at 8{,}192 shots per execution (Table~\ref{tab:cost}).
Validating one layout across multiple windows accumulates shots over the observation period; the two layouts under long-term observation together require approximately 3.7 million hardware shots under the per-run costs and 17/18-window counts in Table~\ref{tab:cost}.
Once a pattern is stored, applying it at compile time is a string-matching scan over the circuit followed by a commuting-gate swap.
Both operations complete in under one second on a single CPU core with no quantum-hardware cost.
In RQ4, successful runs use 11--15 executions (Table~\ref{tab:xalg-runs}). At about 4\,s of metered QPU time per 8{,}192-shot execution, each uses roughly one QPU-minute.

\noindent \textbf{Pattern reuse across compilations.}
The offline cost is amortized across compilations that contain a validated pattern on its target backend and qubits.
Periodic re-validation checks whether the pattern remains active as hardware behavior changes.

\begin{table}[!t]
  \centering
  \small
  \caption{Cost of pattern discovery and use. Offline cost is per single DDMin run; validation multiplies by \# of windows.}
  \label{tab:cost}
   \scalebox{0.8}{
  \begin{tabular}{lcc}
    \toprule
    & \texttt{ibm\_fez} & \texttt{ibm\_marrakesh} \\
    \midrule
    Native gates (one iteration) & 151 & 158 \\
    Segments & 36 & 37 \\
    DDMin convergence steps & 5 & 6 \\
    Oracle calls per run & $\sim$12 & $\sim$14 \\
    Hardware executions per run & $\sim$12 & $\sim$14 \\
    Shots per execution & 8{,}192 & 8{,}192 \\
    Total shots per run & $\sim$98K & $\sim$115K \\
    Validation windows & 17 & 18 \\
    \midrule
    \multicolumn{3}{l}{\emph{Online (per circuit)}} \\
    Pattern scan time & $<$1 s & $<$1 s \\
    Mitigation (commuting swaps) & $<$1 s & $<$1 s \\
    \bottomrule
  \end{tabular}
  }
\end{table}
\section{Discussion}
\label{sec:discussion}

This section discusses the scope and practical implications of our findings.
Section~\ref{sec:threats} examines threats to validity.
Section~\ref{sec:cross-backend} examines cross-backend pattern transfer, while Section~\ref{sec:extensions} discusses limitations and possible extensions.

\subsection{Threats to Validity}
\label{sec:threats}

\noindent \textbf{Internal validity.} Shot noise and calibration drift could produce transient correlations.
We calibrate per-run thresholds against null distributions, shuffle execution order in the controlled intervention, and require persistence across independent calibration windows before accepting a persistent pattern.
When several fragments satisfy the oracle, DDMin's search order may influence which fragment is isolated.

\noindent \textbf{External validity.} We evaluate on IBM Heron backends using 3-qubit Grover circuits for pattern discovery in RQ1--RQ3.
RQ4 extends isolation to three MQT~Bench families (AE, quantum walk, and random circuits) on one backend, two layouts, and one calibration window.
The patterns and their magnitudes may differ on other hardware architectures, qubit topologies, or circuit families.
Cross-window validation is needed to establish whether the additional workloads' isolated fragments recur.

\noindent \textbf{Construct validity.} We use $\mathrm{TVD}(\text{noisy}, \text{real})$ to measure hardware--model mismatch.
To assess sensitivity to this metric, we recompute the ratios for all 192 hardware executions in RQ4 using Hellinger distance.
The resulting ratios correlate strongly with the TVD ratios (Spearman $\rho = 0.874$) and fall on the same side of 1.0 in 99\% of calls (Section~\ref{sec:eval-rq4}).

\subsection{Cross-Backend Pattern Transfer}
\label{sec:cross-backend}

To assess whether a pattern learned on one backend transfers to another with the same topology, we run the \texttt{ibm\_fez} circuit variants unchanged on the same physical qubit indices of \texttt{ibm\_kingston} as a negative control.
The correlation between pattern count and hardware--model mismatch is significant on \texttt{ibm\_fez} but not on \texttt{ibm\_kingston} ($\rho = 0.187$, $p = 0.248$; Figures~\ref{fig:scaling-fez} and~\ref{fig:scaling-kingston}).
These results suggest that shared topology is insufficient to justify reusing execution-derived patterns on another backend.

For compilation, this motivates validating transferred patterns on the target hardware and maintaining backend-specific pattern databases.
The gate sequence can nevertheless serve as a candidate for targeted validation on another backend.
Its source-backend record identifies the sequence and physical neighborhood to test, while measurements on the target backend determine whether the same ordering is associated with hardware--model mismatch there.
This separates reuse of the pattern representation from reuse of its empirical validation: a shared topology supports expressing the same candidate, but does not establish the same execution behavior.
A compiler can therefore retain a common pattern-matching procedure while selecting validated database entries for the backend on which the circuit will execute.

\subsection{Limitations and Extensions}
\label{sec:extensions}

\indent \textbf{Pattern lifetime and re-validation.}
Section~\ref{sec:eval-rq2} shows that the highest-recurrence patterns from the long-term observations were not rediscovered in May--June following the hardware update, while new patterns passed validation.
A production deployment therefore needs periodic re-validation rather than a one-time build.
Re-validation checks whether stored patterns remain useful under the backend\textquotesingle s current operating conditions.

\noindent \textbf{Alternatives to commuting gate swaps.}
Commuting-gate swaps apply to patterns containing an adjacent commuting pair.
Other strategies, such as re-synthesizing the surrounding block, choosing alternative gate decompositions, or changing the routing, could use the same pattern database, but their fidelity gains require evaluation.

\noindent \textbf{Discovery circuit requirements.}
\tool{QRisk}'s pattern discovery uses compact circuits as input.
The input circuit must meet two conditions: (1) its compiled form exercises diverse native-gate combinations on target qubit neighborhoods, and (2) its output distribution lets hardware and reference executions be separated under TVD.
Pattern discovery depends on gate combination coverage, not raw circuit length; that coverage is reachable with compact circuits, and can be extended by adding more circuits rather than scaling any single one up.

\noindent \textbf{Discovery cost and scaling.}
The compact circuits used here keep the measured discovery cost low.
For a circuit of $m$ segments, uncapped DDMin requires $O(m^2)$ oracle evaluations in the worst case~\cite{zeller2002simplifying}.
\tool{QRisk} caps the partition count at $n_{\max}=16$ to limit subdivision.
The 36--37 segment circuits used here required about 12--14 oracle calls (Table~\ref{tab:cost}).
Per-call cost is not constant: each call runs fixed-shot hardware execution plus ideal and noisy simulation, and TVD estimation grows harder as the output space expands with measured qubits.

\noindent \textbf{Physical-qubit coverage.}
Scalability in our current study primarily concerns physical-qubit coverage.
Across three 156-qubit Heron r2 backends, we have exercised 117 of 468 backend-specific physical qubits (25\%).
Extending discovery to additional layouts would increase the physical-qubit coverage of the backend-specific pattern databases.

\section{Related Work}
\label{sec:related}

\noindent \textbf{Delta debugging.} Delta debugging~\cite{zeller2002simplifying} 
automatically isolates failure-inducing inputs through systematic variation.
ProbDD~\cite{wang2021probabilistic} and WDD~\cite{zhou2025wdd} improve delta debugging under probabilistic or weighted assumptions about element relevance.
HDD~\cite{misherghi2006hdd} and Perses~\cite{sun2018perses} exploit input structure (trees, grammars) for faster reduction.
These works improve the search strategy within the DDMin framework.
\tool{QRisk} changes the validation protocol: the output must survive cross-window recurrence validation to account for hardware non-stationarity.
Pontolillo and Mousavi~\cite{pontolillo2024delta} apply delta debugging to property-based regression testing of quantum programs at the Q-SE@ICSE 2024 workshop.
Their oracle uses Fisher's exact tests on measurement distributions to identify source-level changes that reproduce a property-based regression.
\tool{QRisk} instead targets hardware--model mismatch, normalizes each circuit's observed error against a calibration-based reference model, and requires cross-window recurrence before storing a pattern.

\noindent \textbf{Quantum debugging and testing}~\cite{ramalho2025testing,huang2019statistical,muqeet2024qoin}. Sato et al.~\cite{sato2024locating} locate buggy segments in quantum programs by partitioning the source-level circuit and applying statistical testing to identify segments responsible for incorrect output.
Their target is program correctness: the circuit has a specification, and the segment violates it.
\tool{QRisk} targets hardware that produces more error than the compiler's noise model predicts.
The isolated patterns are empirical failure conditions of the backend that the compiler's representation cannot express.
MorphQ~\cite{paltenghi2023morphq} applies metamorphic testing to Qiskit, generating equivalent programs to detect compiler/simulator inconsistencies.
QDiff~\cite{wang2021qdiff} uses differential testing across quantum platforms.
Both techniques test software stack and produce bug reports.
\tool{QRisk} profiles real hardware execution to build reusable pattern database for compilation.

\noindent \textbf{Noise-aware quantum compilation.} Noise-aware compilation uses backend calibration to guide qubit mapping, routing, and gate scheduling~\cite{zhu2025compiler, murali2019noiseadaptive, wagner2025optimized, sharma2023noiseaware, niu2020hardware, huo2025revisiting}.
These approaches optimize compilation using static calibration or characterization signals, including per-gate error rates, $T_1$/$T_2$ times, and measured crosstalk matrices.
\tool{QRisk} complements this information with recurring, backend-specific gate-ordering patterns learned from hardware executions.
The tested gate-ordering effects are not captured by the calibration-based Aer noise model.

\noindent \textbf{Context-dependent error characterization.}
Prior work shows that quantum-circuit outcomes can depend on context variables such as spectator-qubit activity and time, and provides statistical tests for detecting and quantifying crosstalk and drift~\cite{rudinger2019probing}.
Complementary techniques characterize different aspects of device error: cycle benchmarking estimates process fidelity of multiqubit operations in their parallel-execution context~\cite{erhard2019characterizing}; gate-set tomography reconstructs self-consistent gate models and can detect violations of its stationary Markovian assumptions~\cite{blumekohout2017demonstration}; and dedicated crosstalk protocols detect and localize conditional dependencies between subsystems~\cite{sarovar2020detecting}.
\tool{QRisk} is complementary: it uses automated delta debugging to isolate compact circuit fragments associated with hardware--model mismatch without prespecifying a physical error mechanism.
It then validates exact fragments across calibration windows and stores them in a compiler-facing pattern database.

\section{Conclusion}
\label{sec:conclusion}

Compilation mismatch on quantum hardware can be treated as a fault-isolation problem.
\tool{QRisk} adapts delta debugging with a stochastic oracle and circuit-specific threshold calibration to isolate candidate gate patterns associated with hardware--model mismatch.
Cross-window recurrence validation retains independently rediscovered patterns in a backend-specific pattern database.
On MQT~Bench workloads (Section~\ref{sec:eval-rq4}), \tool{QRisk} isolates 3--5-gate fragments in 10 of 12 runs, using 11--15 oracle calls per successful run.
As Section~\ref{sec:eval-rq3} shows, simple commuting-gate swaps guided by the pattern database reduce mean hardware--model mismatch by 24\% and 45\% on the two tested backends.

\section{Data Availability}
\label{sec:data-availability}

The \tool{QRisk} artifact is available on Figshare at
\url{https://figshare.com/xxxxxx}. The artifact is intended to support review and reproducibility.

\bibliographystyle{ACM-Reference-Format}
\bibliography{main}

\end{document}